# The 3D-DCT transform: didactic experiment and possible useful tool

Fernando Martín-Rodríguez [(1)], Fernando Isasi-de-Vicente [(1)], Mónica Fernández Barciela [(1)].
fmartin@uvigo.es, fisasi@uvigo.es, monica.barciela@uvigo.es.
[(1)] AtlantTIC research center for Telecommunication technologies. Universidad de Vigo.
E.E.T. C/Maxwell SN (Ciudad Universitaria), 36310 Vigo (Pontevedra).

*Abstract*- **This article is about an image transform called 3D-DCT, or three-dimensional discrete cosine transform. This is an extension of the well-known 1D and 2D-DCT, which is extensively used, mostly in multimedia coding. A modification of 1D-DCT (MDCT) is used in audio coding, whereas 2D-DCT is the key concept for JPEG standard (still image coding) and also for very popular video coding standards: MPEG-1 and 2, H.261 & H.263. Other, more modern methods like MPEG-4/H.264 and the newest MPEG-H/H.265 use a modified version of 2D-DCT. Nevertheless, 3D-DCT is less known. In this work, a simple implementation for 3D-DCT is constructed and some applications are explored. This is designed as a material for class work but, as it will be shown in this paper, some practical applications can be interesting. 3D-DCT can be applied for creating a video coder much simpler than those following MPEG-x/H.26x and with only slightly less performance. 3D-DCT has also been tested for 3D medical imaging (CT studies) coding, yielding very interesting results.**

## I. Introduction

Let be a monochrome square image. This image is mathematically represented as a discrete matrix or function of two integer variables: **x[m,n]** (**m,n=0**...N-1). Under these conditions, the two-dimensional discrete cosine transform (2D-DCT) of **x** is defined as (**$S_x$**):

$$S_x[u,v] = \alpha(u)\alpha(v) \sum_{m,n=0}^{N-1} x[m,n] \cos\left(\frac{\pi}{2N}(2m+1)u\right) \cos\left(\frac{\pi}{2N}(2n+1)v\right) \quad (1)$$

Where: $\alpha(x) = \begin{cases} 1/\sqrt{N}, & x = 0 \\ \sqrt{2/N}, & x \neq 0 \end{cases}$

This definition, as can be deduced from the formula, allows obtaining a new real matrix (no imaginary part) of the same size (NxN). It can be shown that, when applied to low-pass spectrum images (all natural images and those synthetic images that look natural), most of the information is concentrated in the first coefficients: the **$S_x$ [u,v]** with **u+v** of small value. This fact is simply due to the concentration of information at low frequencies. The **u** and **v** indices can be understood as horizontal and vertical spatial frequencies.

The qualitative properties of the DCT are very similar to those of the Fourier transform. An FFT-2D would also concentrate the information at low frequency but, in this case, we would have a non-zero imaginary part and a replication of the transform values at multiples of N (periodicity in the two spatial frequencies with period $2\pi$).

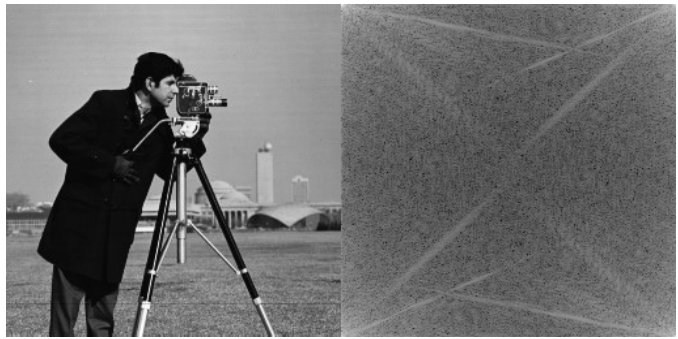

Fig. 1. Image (left) and modulus (logarithm) of its FFT-2D (right).

Of course, for DCT, there is a reversal formula that allows the original image to be recovered **exactly**:

$$x[m,n] = \sum_{u,v=0}^{N-1} S_x[u,v]\alpha(u)\alpha(v) \cos\left(\frac{\pi}{2N}(2m+1)u\right) \cos\left(\frac{\pi}{2N}(2n+1)v\right) \quad (2)$$

Widely used methods, such as JPEG, MPEG-1 and 2 standards, H.261 and H.263 still images, are based on dividing the images into small square blocks (the size used in these standards is 8x8), calculating the 2D-DCT transform of each block and quantifying the S coefficients $S_x[u,v]$.

This work consists in studying the extension to three dimensions of the DCT concept. Id EST: starting from a three-dimensional (cubic) matrix of real values: **x[m,n,p]** (m,n,p=0...N-1), the 3D-DCT is defined as (analysis equation):

$$T_x[u,v,w] = \alpha(u)\alpha(v)\alpha(w) \sum_{m,n,p=0}^{N-1} x[m,n,p] \cos\left(\frac{\pi}{2N}(2m+1)u\right) \cos\left(\frac{\pi}{2N}(2n+1)v\right) \cos\left(\frac{\pi}{2N}(2p+1)w\right) \quad (3)$$

And the inversion formula (or synthesis equation):

$$x[m,n,p] = \sum_{u,v,w=0}^{N-1} T_x[u,v,w]\alpha(u)\alpha(v)\alpha(w) \cos\left(\frac{\pi}{2N}(2m+1)u\right) \cos\left(\frac{\pi}{2N}(2n+1)v\right) \cos\left(\frac{\pi}{2N}(2p+1)w\right) \quad (4)$$

This concept has already been formulated, although it is not very well known. It has been used for video coding [1] and also for coding some types of medical images [2].

## II. METHODS AND MATERIALS

### A. Starting point

Our work is conceived as a teaching experiment in which students are expected to carry out a small practical work on coding. We start by providing them with a functional MATLAB code [3] that implements still image JPEG encoding (Figure 2).

```matlab
function ImCod = CodJPEG(im0,Q)
%%%%%%%%%%%%%%%%%%%%%%%
% Codificacion JPEG %
%%%%%%%%%%%%%%%%%%%%%%%
NB = length(Q);
[filas,columnas] = size(im0);
%
% Codificar por bloques
k = 0;
for fila=1:NB:filas
    for columna=1:NB:columnas
        bloque = im0(fila:fila+NB-1,columna:columna+NB-1);
        bloquedct = dct2(bloque);
        bloqueq = round(bloquedct./Q);
        bloques = ZigZag2(bloqueq);
        bloquebin = HuffmanCod(bloques);
        k = k+1;
        ImCod(k) = bloquebin; %#ok<AGROW>
    end
end
```

Fig. 2. Matlab code implementing JPEG with 2D-DCT.

In the code, we can see how the image is divided into small square blocks of the same size as the quantization matrix: Q. Normally, Q is of size 8x8 but the present code can work with any square size.

The cosine transform in two dimensions is applied to each of the original blocks (line 13). The values of the cosine transform are quantized. The matrix Q contains a different quantization step: $Q_{ij}$ for each DCT coefficient, according to its position or importance). The next step is the zig-zag sorting of the coefficients in order of importance (line 14). Note that, to generate a sequence of data that can be saved in a file, it will always be necessary to serialize the matrices (reorder them as vectors). The zig-zag serialization (Figure 3) places the least significant coefficients at the end. This, for natural images, produces long series of zeros at the end of the quantized and serialized block (because the high-frequency coefficients are weak and are quantized more strongly). These series of zeros (in a typical JPEG encoding there are more than 90% of zeros at this stage) allow a high compression of the data, using an end-of-block binary code in the final binary sequence. In line 16, a variable length binary encoding (Huffman code) is applied, which will be responsible for the reduction in the size of the information. In addition to introducing the end-of-block code at the time when the remaining coefficients are a series of zeros, the significant part is also encoded using shorter codes for more probable values and vice versa.

The coefficients of the Q matrix were calculated by A.B. Watson in a popular paper [4], and the values are calculated to minimize the visual effect. The original matrix has these values:

$$Q = \begin{bmatrix} 016 & 011 & 010 & 016 & 024 & 040 & 051 & 061 \\ 012 & 012 & 014 & 019 & 026 & 058 & 060 & 055 \\ 014 & 013 & 016 & 024 & 040 & 057 & 069 & 056 \\ 014 & 017 & 022 & 029 & 051 & 087 & 080 & 062 \\ 018 & 022 & 037 & 056 & 068 & 109 & 103 & 077 \\ 024 & 035 & 055 & 064 & 081 & 104 & 113 & 092 \\ 049 & 064 & 078 & 087 & 103 & 121 & 120 & 101 \\ 072 & 092 & 095 & 098 & 112 & 100 & 103 & 099 \end{bmatrix} \qquad (5)$$

This matrix is used to achieve an average quality (q=50%). To achieve higher qualities (50<q<100) it is multiplied by a factor 2-q/50, always less than 1, achieving smaller quantification steps. For q<50 the factor is 50/q, greater than 1, which increases the quantification steps.

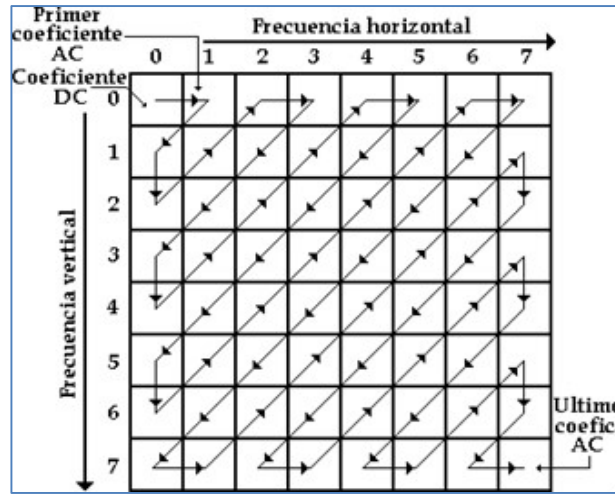

Fig. 3. Zig-zag ordering from JPEG standard.

The exercise consists, in addition to studying and evaluating this encoder, of converting it into a video encoder using the concept of 3D DCT and stacking consecutive frames of the video in order to obtain cubic three-dimensional blocks of size 8x8x8.

It is proposed to the students evaluate this JPEG encoder with a collection of fourteen uncompressed grayscale images (bmp format) including some of the most historically used (Lenna, Cameraman). The results obtained for the Q matrix of (5), q=50, are PSNR 34 dB and compression degree 9.40 (size-original/size-compressed).

*B. 3D-DCT computation (by separability)*

To implement the calculation of the three-dimensional DCT, we need to develop a new function, since there is no such extension in the MATLAB catalog.

For this purpose, an advantage is taken of the separability property. That is, a 2D transform is separable if it can be written in this way:

$$S_{NxN} = C_{NxN} x X_{NxN} x C_{NxN}^T \qquad (6)$$

Where X is the input image (matrix), S is the transform (matrix) and C is the matrix that defines the transformation. The first product in 6, would be the 1-D transform of the columns of X, while the second would be the 1-D transform of the rows. That is: the 2-D transform can be calculated through 1-D transforms: it can be done first by rows and then by columns or vice versa, and the result is the same.

The DCT is separable, in terms of 6, we would have to do: $c_{ij} = \alpha(i-1)\cos(\frac{\pi}{N}(2j-1)(i-1))$, with i and j in a range between 1 and N.

Applying this to a three-dimensional case, given an input cube **x[m,n,p]**, the DCT-3D can be calculated:
- First, apply the DCT-2D to each 2D layer of **x**: **x[m,n,p₀]** (**p₀** constant).
- Applying in the resulting matrix, the DCT-1D (**dct** function in MATLAB), to each vector **T₂d [m₀,n₀,p]** (**m₀** and **n₀** constant).

*C. From the quantization matrix to the quantization cube*

Now it is necessary to calculate a three-dimensional quantization matrix. That is a quantization cube. Calling this cube Q3, it will be a matrix with three indices Q3(i,j,k) where the indices i,j, and k indicate spatial frequencies and, intuitively, the values of Q3 should grow as the sum i+j+k increases.

To easily calculate an operational matrix, the following method has been designed:

Make the matrices corresponding to the Cartesian planes, that is Q3(i,j,1), Q3(i,1,k), and Q3(1,j,k) equal to the original Watson matrix (equation 6).

Calculate the remaining values according to the plane $i+j+k=c_0$ to which they belong.

For a plane, $i+j+k=c_0$, make its coefficients Q(i,j,k) not yet assigned; equal to the average of those already assigned (the values already assigned are those on the intersection lines of that plane with the Cartesian planes).

For $c_0 \geq 18$, there is no intersection. For these high frequencies, we make Q3(i,j,k) = 100.

Note that the DCT-3D will have a different scaling. According to formula 1, the maximum value of a DCT-2D is **max_lum x N** (continuum coefficient of a constant block equal to the maximum luminance value). For the typical situation: **max_lum=255** and **N=8**, a maximum value of 2040 is obtained. In DCT-3D (formula 3), the maximum will be **max_lum x N$^{1.5}$**. For the same case, a maximum of 5770 is obtained. This data seems to suggest that one could scale the Q3 matrix by a factor of 5770/2040 = 2.83, maintaining the quality/compression ratio. This is an estimate; obviously, it must be tested with real videos since the temporal redundancy in the third dimension is not going to be equal to the spatial redundancy. Experiments have been performed with the Q3 matrix derived from the original Q, without scaling, obtaining satisfactory results: see section 3.

*D. Serialization of the quantized block*

Serialization is simply a way of ordering the coefficients of the transformed matrix already quantized. It should be an ordering by perceptual importance, or in other words: an ordering that favors the appearance of long series of zeros, especially at the end of the block, so that variable-length coding (lossless compression) is very efficient.

To perform the 3D transform path, it was experimentally verified that 90% of the non-zero coefficients of the block are in the first layer: **T$_{3d}$[i,j,1]**, where the third dimension is the temporal one. This is undoubtedly due to the higher temporal redundancy compared to the spatial one.

Therefore, the serialization is performed in a simple way: by layers: **T$_{3d}$[i,j,k$_0$]**, with constant **k$_0$**. In each layer, the zig-zag path inherited from the standard JPEG method is applied

*E. Variable length encoding*

This section does not change compared to the 2D view. However, we will comment a little on its implementation.

For this purpose, the tables published in [5] are used. It should be noted that, actually, the numerical values are coded in one's complement and the Huffman code expresses the length of the intermediate series of zeros (if any) and the length of the code word in one's complement (the most significant bit lets us know the sign of the coefficient).

Given figure 4, the process can be explained with a couple of examples:
- All blocks start with a DC coefficient. There is no series of leading zeros. A binary sequence '110' at the beginning of the block would mean that we must read the next 5 bits to get the DC coefficient.
- If the next coefficient were equal to '-10', it would be represented as '0101' in one's complement. We would find: '1011' in the sequence, meaning: no intermediate zeros, read the next 4 bits. Those next 4 bits would be: '0101', the '0' in the most significant position indicates a negative sign.
- If we had a zero coefficient followed by a '+3' value, the Huffman code would be '11011' followed by '11' (a series of only one zero followed by a two-bit code that corresponds to '+3' in one's complement).
- The sequence '1010' or "end of block (EOB)" indicates that the remaining coefficients are all zero, moving on to the next block.

Fig. 4. Part of the Huffman tables used.

*F. Video encoding with 3D-DCT*

With all these ingredients, monochrome video coding can be implemented with a simple code very similar to the original one in Figure 1. In Figure 3, we can see how the video is represented with a three-index matrix that is divided into cubes to which the same process is applied: transformation, quantization, serialization, and Huffman coding.

```matlab
function [VidCod,filas,columnas] = CodJPEGvid(Vid,Q3)
%%%%%%%%%%%%%%%%%%%%%%%%%%%%%%%%%%%%%%%%
% Codificacion de video mediante DCT 3D %
%%%%%%%%%%%%%%%%%%%%%%%%%%%%%%%%%%%%%%%%
NB = length(Q3);
[filas,columnas,L] = size(Vid);
%
% Codificar por cubos NBxNBxNB.
% gop = (group of pictures): NB consecutive frames
k = 1;
for gop=1:NB:L
    i = 1; % Incializar gop
    for fila=1:NB:filas
        for columna=1:NB:columnas
            cubo = Vid(fila:fila+NB-1,columna:columna+NB-1,gop:gop+NB-1);
            cubodct = DCT3D(cubo);
            cuboq = round(cubodct./Q3);
            cubos = Serializar3D(cuboq);
            cubodef = HuffmanCod(cubos);
            VidCod(i,k) = cubodef;          %#ok<AGROW>
            i = i+1;
        end
    end
    k = k+1;
end
```

Fig. 5. Matlab code implementing video coding with 3D-DCT.

*G. Test Videos*

Twenty-three test videos will be used. All of them are color videos in **.y4m** format (without compression). Among them are some very popular in MPEG testing: Akiyo, Foreman, Claire, Miss America... A function obtained from MATLAB Central [6] is used to read this format. This function returns a **movie** object from which the Y, $C_b$, $C_r$ components required for encoding can be easily extracted (for monochrome video the Y component is enough).

III. RESULTS AND DISCUSSION

*A. Monochrome video encoding*

The first test scenario was that of monochrome (grayscale) video encoding. For the first video (Akiyo, 30 fps, CIF format: 352x88) a PSNR of 41 dB and a bit rate of 0.47 Mb/s is obtained. The uncompressed bit rate would be 24.30 Mb/s. The compression ratio (speed quotient) would amount to 51.20. This result is obtained for the Q3 matrix derived from the original Watson Q (q=50 in JPEG).

Encoding the same video with MPEG-4 (using **videowriter** in MATLAB with default options); results in a PSNR of 45 dB and a bit-rate of 0.42 Mb/s. It is undoubtedly better but there is not as much difference as could be expected: 10% more in PSNR and 11% less in bitrate. As 3D-DCT is a much simpler method, it could be an option for encoders that need to run on simple devices or even in cases where there is an interest in editing the video before distributing it. When this need exists (especially in video production companies), formats without temporal redundancy such as DV-25, DVC-pro... are currently used, which are much less efficient.

Obtaining average data for the complete video collection, the results are 37.50 dB PSNR, 1.17 Mb/s bitrate, and an average compression ratio of 29.68. For MPEG-4 we obtain 42.80 dB PSNR (14% more) and 1.07 Mb/s (8% less).

*B. Color video encoding*

The color video is encoded in the same way as monochrome video: by running the encoding three times, once for each component (Y, $C_b$, and $C_r$). The chroma components are under-sampled with a 4:2:0 scheme (dividing both the width and height of each frame by 2), as in MPEG. In addition, a different matrix is used for color coding ($Q3_c$), obtained from the $Q_c$ matrix of the 2-D JPEG (which has slightly larger quantization steps, see equation 7).

$$Q_c = \begin{bmatrix} 017 & 018 & 024 & 047 & 099 & 099 & 099 & 099 \\ 018 & 021 & 026 & 066 & 099 & 099 & 099 & 099 \\ 024 & 026 & 056 & 099 & 099 & 099 & 099 & 099 \\ 047 & 066 & 099 & 099 & 099 & 099 & 099 & 099 \\ 099 & 099 & 099 & 099 & 099 & 099 & 099 & 099 \\ 099 & 099 & 099 & 099 & 099 & 099 & 099 & 099 \\ 099 & 099 & 099 & 099 & 099 & 099 & 099 & 099 \\ 099 & 099 & 099 & 099 & 099 & 099 & 099 & 099 \end{bmatrix} \qquad (7)$$

The average results are now: 33.33 dB PSNR, 1.26 Mb/s bitrate, and an average compression ratio of 78. For MPEG-4 we obtain 35.90 dB PSNR (8% more) and 1.13 Mb/s (10% less).

The conclusion is the same, although slightly inferior in performance to MPEG, it can be an alternative in special environments.

*C. Medical images coding*

In [2], the authors use 3D-DCT for the encoding of color video from endoscopies. Here a different application is tested: the encoding of 3D medical studies. A 3D medical study (best examples are computed tomography, CT, or magnetic resonance imaging, MRI) consists of a set of monochrome images corresponding to parallel slices of areas of the human body.

These studies are usually stored in collections of files stored in a specific format for the work and exchange of medical images: the DICOM format.

However, when they are read by an application, an in-memory structure consisting of a numerical array with three indices is usually created. That is: it is very voluminous information that can be efficiently encoded using 3D-DCT.

Of course, any 3D-DCT-based format is not going to replace DICOM which will continue to be used on a day-to-day basis. However, some kind of compression may be interesting to facilitate fast file exchange and/or for massive archiving of old medical records.

Testing with a real CT of 460 slices, each of size 512x512 (12-bit monochrome images) yielded: 47.43 dB PSNR and a compression ratio of 116.51. The quality seems sufficient but could be increased by using a Q3 matrix scaled by a number less than 1. The original DICOM files use a lossless type of compression (JPEG-2000 lossless). If the degree of compression is measured by dividing the total size of these files by the compressed 3D-DCT size, the factor obtained is 22.

## IV. Conclusions and Future Work

This paper explores the possibilities of the three-dimensional cosine transform (3D-DCT), considering a teaching exercise but also measuring its potential for video coding as well as for medical image compression.

In the future, it will be possible to further explore potential applications, especially in the field of 3D medical image coding.

## Acknowledgments

The authors would like to thank the University of Vigo and the AtlantTIC center for their support in this work.